# Electrical spin injection in p-type Si using Fe/MgO contacts

A. Spiesser[*a], S. Sharma[a,b], H. Saito[a], R. Jansen[a], S. Yuasa[a], and K. Ando[a]
[a]National Institute of Advanced Industrial Science and Technology (AIST),
Spintronics Research Center, Tsukuba, Ibaraki 305-8568, Japan
[b]Zernike Institute for Advanced Materials, University of Groningen, The Netherlands

**ABSTRACT**

We report the successful electrical creation of spin polarization in p-type Si at room temperature by using an epitaxial MgO(001) tunnel barrier and Fe(001) electrode. Reflection high-energy electron diffraction observations revealed that epitaxial Fe/MgO(001) tunnel contacts can be grown on a (2×1) reconstructed Si surface whereas tunnel contacts grown on the (1×1) Si surface were polycrystalline. Transmission electron microscopy images showed a more flat interface for the epitaxial Fe/MgO/Si compared to that of the polycrystalline structure. For the Fe/MgO/p-Si devices, the Hanle and inverted Hanle effects were clearly observed at 300 K by using a three-terminal configuration, proving that spin polarization can be induced in the Si at room temperature. Effective spin lifetimes deduced from the width of the Hanle curve were 95 ± 6 ps and 143 ± 10 ps for the samples with polycrystalline and epitaxial MgO tunnel contacts, respectively. The observed difference can be qualitatively explained by the local magnetic field induced by the larger roughness of the interface of the polycrystalline sample. The sample with epitaxial Fe/MgO tunnel contact showed higher magnitude of the spin accumulation with a nearly symmetric behavior with respect to the bias polarity whereas that of the polycrystalline MgO sample exhibited a quite asymmetric evolution. This might be attributed to the higher degree of spin polarization of the epitaxial Fe/MgO(001) tunnel contact, which acts as a spin filter. Our experimental results suggest that an epitaxial MgO barrier is beneficial for creating spins in Si.

**Keywords:** spintronics, electrical spin injection, spin accumulation, silicon, epitaxial Fe/MgO tunnel contact, Hanle measurement, spin lifetime, bias dependence.

## 1. INTRODUCTION

In existing electronics, information storage is realized in magnetic media whereas information processing is done with silicon-based devices. For more than a decade, the idea of having a device that can combine both functions has been the driving force in spin-based semiconductor research. The impact of this field is now widely recognized as it should lead to the development of a new generation of devices having non-volatile memory functionality, high processing speed, high integration density and low power consumption. One of the key requirements for the development of such devices is the ability to create spin-polarized carriers in a semiconductor (SC) by electrical means. Silicon is an ideal host SC material because of its compatibility with the existing complementary metal-oxide semiconductor (CMOS) technology. Recently, the electrical creation of spin polarization has been achieved in Si at room temperature (RT) by using tunnel contacts[1-7]. These results open the way towards integration of spintronics into conventional Si technology, and the creation of novel functionalities.

The fabrication of high-quality ferromagnet (FM)/oxide tunnel contacts on a SC is a prerequisite to achieve effective creation of spin polarization in the SC. Magnesium oxide (MgO) is the canonical tunnel barrier material for metal-based spintronics devices[8-10]. It is well known that a single-crystalline MgO(001) barrier acts as a spin filter by selecting states with well-defined symmetries that give rise to high tunnel spin polarization, and thus, giant tunneling magnetoresistance in MgO-based magnetic tunnel junctions (MTJs)[8-10]. Hence, one can expect that MgO is also an effective tunnel barrier for Si-based spintronics devices if a thin MgO(001) layer can be epitaxially grown on Si. Whereas epitaxial growth of FM/MgO(001) for spin injection on Ge[11-17] and GaAs[18-20] has been intensively studied, the way towards epitaxial growth of MgO on Si is still not well established. Therefore, it is important to clarify the growth techniques and conditions for achieving epitaxial FM/MgO(001) layers on Si[4,7,21-23]. In addition, the relationship between the crystal quality of the MgO barrier and the important spin-related physical parameters in Si, such as the spin lifetime and the magnitude of spin accumulation signal, should be investigated.

[*]aurelie.spiesser@aist.go.jp; Phone: +81-29-861-8382; Fax: +81-29-861-3432

Here, we report the fabrication of Fe/MgO tunnel contacts on heavily doped p-type Si (p-Si) and demonstrate electrical creation of spin polarization at RT by employing Hanle measurements in a three-terminal configuration. Depending on the surface condition of the Si, polycrystalline or epitaxial Fe/MgO tunnel contacts can be grown. We demonstrate that the spin lifetime and the magnitude of spin accumulation signal are larger for samples with epitaxial MgO tunnel barrier than for the polycrystalline samples. Our experimental findings lead to a better understanding of the effect of the crystal quality of the tunnel barrier on the spin accumulation signal and show that epitaxial MgO is desirable for achieving efficient spin injection in Si.

## 2. EXPERIMENTAL DETAILS

The Fe/MgO tunnel contacts were grown by molecular beam epitaxy (MBE) on p-type Si(001) substrates with a hole concentration ($p$) of $4.8 \times 10^{18}$ cm$^{-3}$ at 300 K (dopant: boron). Before loading the substrates into the ultra-high vacuum chamber, a chemical cleaning in acetone and isopropanol was carried out to eliminate hydrocarbon related contaminants. Then, the Si wafers were etched in a buffered HF solution, which leads to a hydrogen-terminated Si surface. The substrates were introduced in the MBE system with a base pressure less than $5 \times 10^{-10}$ Torr. In this study, we have grown the Fe/MgO tunnel contacts on Si substrates with two different surface preparation conditions. The first one is a non-annealed Si substrate (sample A), for the other one the Si was annealed at 700°C for 10 min (sample B). For both samples, the MgO tunnel barrier (2 nm) and Fe electrode (10 nm) were deposited by electron-beam evaporation at 300°C and 100°C, respectively. The former is the optimum temperature for the growth of epitaxial MgO on Si(001)[22]. Finally, a 20-nm-thick Au capping layer was evaporated by using a conventional Knudsen-cell.

Structural analyses were performed by *in-situ* reflection high-energy electron diffraction (RHEED) and high-resolution transmission electron microscopy (HRTEM). For the electrical transport measurements, junctions with an active tunnel area ($A$) of 100×200 µm$^2$ were prepared by standard microfabrication techniques (lithography, Ar-milling, SiO$_2$ sputtering and lift-off). Current-voltage characteristics were measured with a conventional two-probe method. For detection of the induced spin accumulation, Hanle measurements in a three-terminal configuration were performed at 300 K in a superconducting magnet equipped with a sample rotator.

## 3. RESULTS AND DISCUSSION

### 3.1 Structural characterizations

We found that the crystal quality of the Fe/MgO layers is quite sensitive to the surface condition of the Si substrate before deposition. Figure 1 presents the evolution of the RHEED patterns for the samples A and B before and after the deposition of the MgO and the Fe layers. The surface of the non-annealed Si (sample A) displayed a (1×1) RHEED pattern, as shown in Fig. 1(a). The pattern turned into faint rings after the deposition of the MgO layer (Fig. 1b), indicating that the MgO is not single crystalline. The pattern of the Fe layer on the MgO showed broad spots corresponding to Fe(001) with faint ring patterns (Fig. 1c), suggesting a coexistence of (001)-oriented and polycrystalline Fe. In contrast, if the Si substrate is annealed at 700°C (sample B), additional streaks appeared between the main (1×1) streaks along the [110] azimuth (Fig. 1d). The MgO layer grown on this well-ordered (2×1) reconstructed Si surface revealed spotty patterns along the [110] azimuth corresponding to MgO(001) (Fig. 1f). Finally, after Fe deposition, a spotty (1×1) RHEED patterns is observed along [100] azimuth, but no rings were visible (Fig. 1g), indicating a single crystalline Fe(001). Consequently, epitaxial Fe(001)/MgO(001) structures can be successfully grown on Si(001). These results clearly indicate that the (2×1) Si surface is crucial for the fabrication of epitaxial MgO(001) on Si. Mönch *et al.* have reported that the surface energy of the (2×1) Si surface is lower than that of the (1×1) surface[19]. This might be an important parameter for this epitaxial growth mechanism.

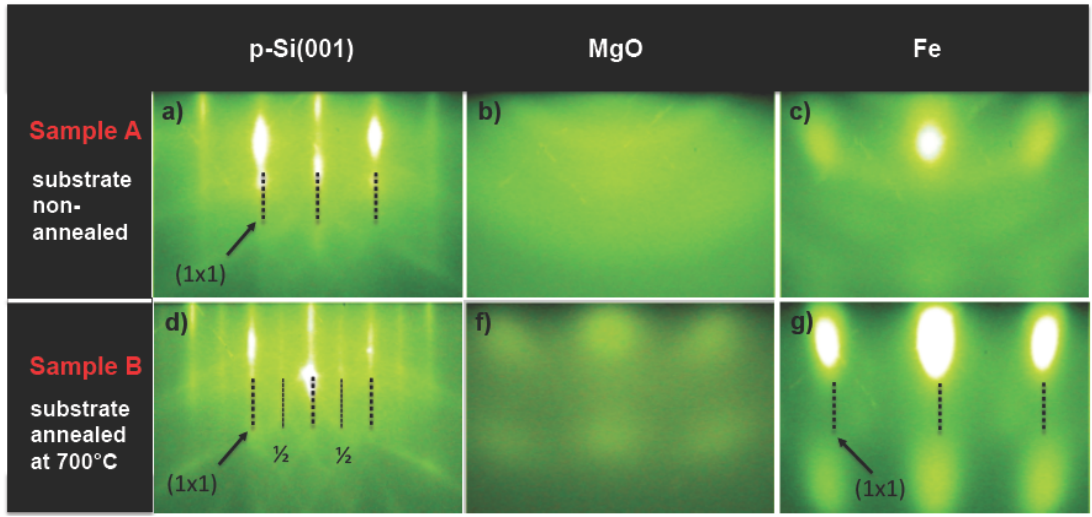

Figure 1 RHEED patterns of Si, MgO and Fe layers of the samples A (upper panels) and B (lower panels); (a) non-annealed Si(001) surface (b) MgO and (c) Fe layers of sample A, (d) Si(001) surface after annealing at 700°C for 10 min, (f) MgO and (g) Fe layers of sample B. Bulk (1×1) and (2×1) reconstructed patterns are indicated by thick and thin dashed lines, respectively.

We have performed TEM observations for further investigation of the crystal quality of the two samples. Figure 2 displays the cross-sectional TEM images of the samples A and B taken along the [110] direction of the Si substrate. Basically, both samples have sharp interfaces without interdiffusion and/or intermixing between Si, MgO and Fe layers. As expected from the RHEED observations, there is no crystallographic relationship between the Si, MgO and Fe layers in sample A, as shown in Fig. 2 (a). In the MgO layer, we can observe different crystal orientations separated by amorphous zones. The interface with Fe is quite rough (about 0.3 nm) and the Fe layer exhibits a fully polycrystalline structure. On the contrary, a unique orientation of the MgO crystal is observed in the sample B, with slightly tilted planes, as shown in Fig. 2 (b). The Fe/MgO interface is smoother than in sample A (about 0.1 nm) and the Fe layer is clearly single crystalline. Another noticeable observation concerns the MgO/Si interface: the roughness of sample A is larger (3 to 4 monolayers) than that of the sample B (about 2 monolayers). By combining the RHEED results and the TEM analysis, we confirm the following epitaxial relationship in sample B: Fe(001)/MgO(001)/Si(001) and Fe[100]//MgO[110]//Si[110], which is in agreement with the previous work[4,7,22].

To visualize the crystallographic relationship of sample B, we show in Fig. 2(c) schematic images of the Si, MgO and Fe lattice, with top and perspective views. The MgO(001) grows on Si(001) with a cube-on-cube relationship despite the large lattice mismatch between Si and MgO (the lattice constant of Si is 5.431 Å and that of MgO is 4.211 Å, corresponding to a direct lattice mismatch of 22.5 %). Note that 4 MgO unit cells almost match with 3 Si unit cells, resulting in a much smaller lattice mismatch of 3.9%. This could be an effective mechanism for the cube-on-cube relationship between the MgO(001) and Si(001). The top Fe cell is rotated in the plane by 45° with respect to the MgO and Si cells. From these results, we might expect a higher tunneling spin polarization[8-10] of sample B with the epitaxial tunnel contact. Also, further improvement of the crystalline quality of the MgO barrier layer can be achieved by optimizing the Si surface preparation conditions.

## 3.2 Spin accumulation in p-type Si using Fe/MgO contacts

First, we measured the current-voltage (*I-V*) characteristics for sample A and B at RT, as shown in Fig. 3 (a). The bias voltage is defined as $V_{Si}$ - $V_{Fe}$, where $V_{Si}$ and $V_{Fe}$ are the potential of the Si and Fe, respectively. Both samples exhibit a similar nonlinear and nearly symmetric behavior with respect to the bias polarity. We could not observe the strongly rectifying behavior that is typical of a conventional Schottky diode, indicating that tunneling through the MgO is the dominant electrical transport mechanism[25].

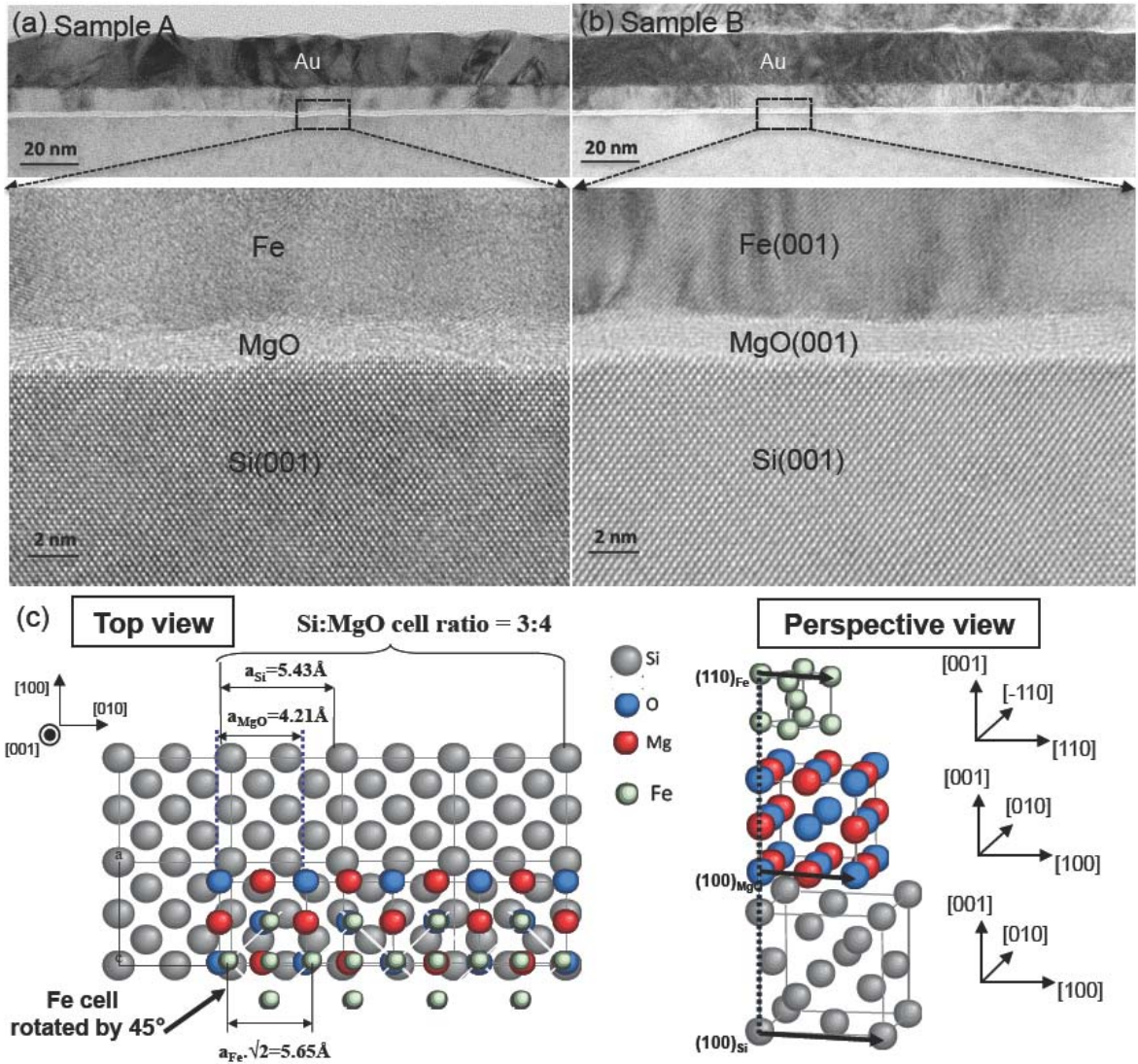

Figures 2 (a) and (b) High-resolution cross-sectional TEM images of the samples A and B, respectively, (c) Schematic illustrations of the epitaxial relationship for Fe(001)/MgO(001)/Si(001) with Fe[100]//MgO[110]//Si[110] in the sample B from top and perspective views.

To detect the spin accumulation in Si, we performed so-called Hanle measurement in a three-terminal configuration[1,5 26-28], as shown in Fig. 3 (b). The central FM tunnel contact 2 is the spin injector/detector, and the contacts 1 and 3 are the reference electrodes. The dimensions of the FM contacts (100×200 μm) and the distances between the contacts (500 μm) are designed to be much larger than the spin-diffusion length of Si. This guarantees that the spin accumulation occurs just underneath the FM contact without any spin-related interference among the contacts.

For a conventional Hanle effect measurement, a magnetic field $B_\perp$ is applied perpendicular ($\perp$) to the film plane, while the FM electrode is magnetized parallel ($\parallel$) to the film plane. This causes precession of the spins in the Si and a reduction of the spin accumulation. At constant tunnel current, this results in a reduction of the tunnel voltage with a Lorentzian line shape. However, Dash et al. has recently demonstrated that the spin accumulation can be partially suppressed by precession in inhomogeneous local magnetic fields $B_{loc}$ arising from the finite roughness of the FM[5]. This effect can be suppressed by applying an in-plane field ($B_\parallel$), leading to an increase in tunnel resistance due to the recovery of spin accumulation (inverted Hanle effect[5]). Then, the magnitude of the spin accumulation signal ($\Delta V_{spin}$) corresponds to the sum of the voltage change in the Lorentzian part of the Hanle and inverted Hanle curves.

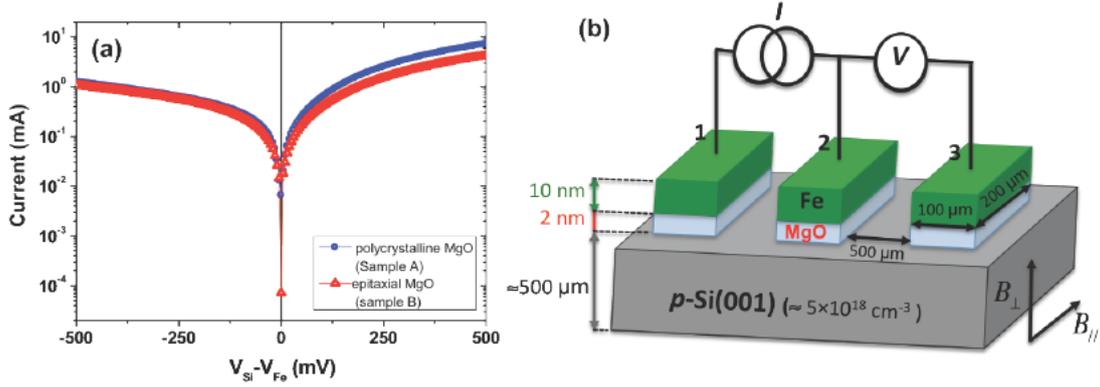

Figures 3 (a) Current-voltage characteristics of the Fe/MgO/p-Si devices with polycrystalline MgO (sample A, blue circles) and epitaxial MgO (sample B, red triangles) barriers measured at RT by a conventional two-probe method. The bias voltage is defined as $V_{Si} - V_{Fe}$, where $V_{Si}$ and $V_{Fe}$ are the potentials of the Si and Fe electrodes, respectively, (b) Schematic device structure with three-terminal configuration for Hanle ($B_\perp$) and inverted Hanle ($B_{//}$) measurements.

The voltage changes $\Delta V$ for the samples under $B_\perp$ and $B_\parallel$ at 300 K are presented in Figs. 4. Positive and negative bias correspond to hole extraction and hole injection into the Si valence band, respectively. Clear Hanle and inverted Hanle signals were observed for the two samples at both polarities, demonstrating electrically induced spin polarization in p-Si at RT. Note that the observation of sizable inverted Hanle signals indicates that the induced spins in the Si are strongly affected by the $B_{loc}$.

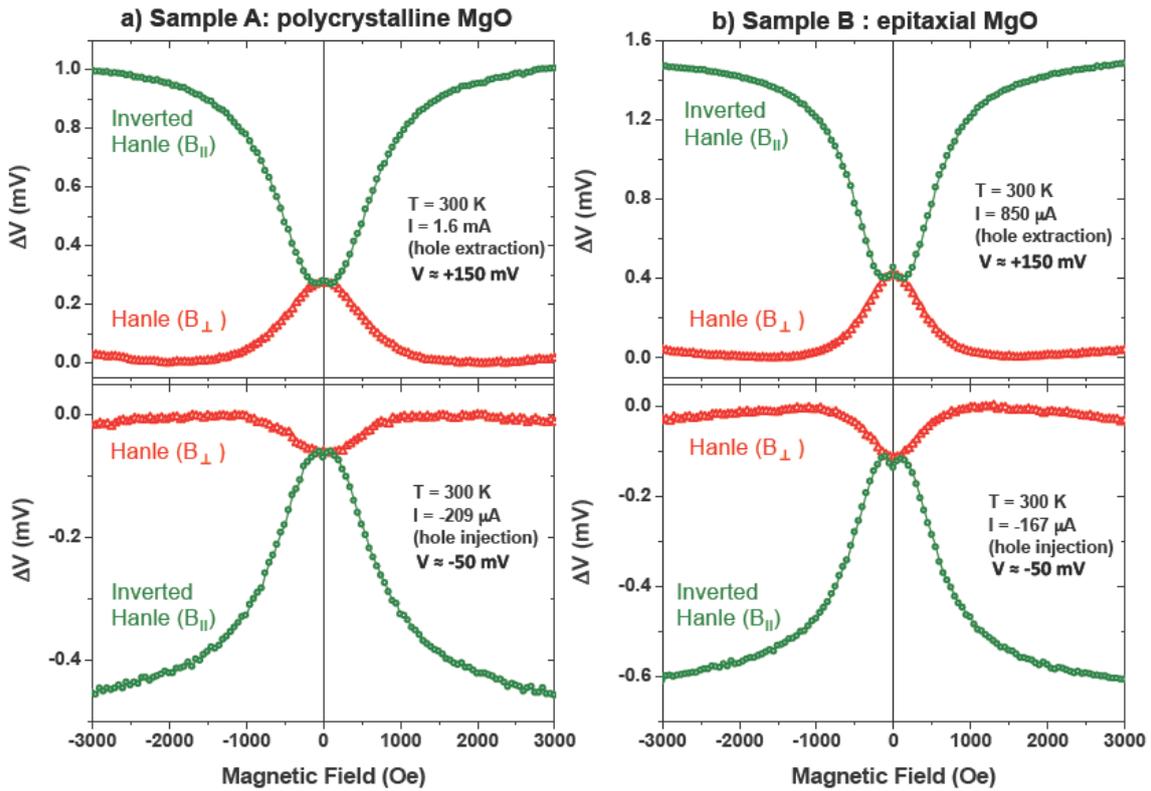

Figures 4 Hanle ($B_\perp$, red open triangles) and inverted Hanle ($B_{//}$, green open circles) curves for the Fe/MgO/p-Si devices with (a) polycrystalline Fe/MgO (sample A, left panels) and (b) epitaxial Fe/MgO (sample B, right panels) barriers. All measurements were performed at 300 K.

From the observed Hanle curves, we estimated the spin lifetime $\tau_s$ in the p-Si. Figure 5 displays the Hanle curves of both samples measured at 300 K with negative and positive bias voltages. When $B_\perp$ is small, the decay of the voltage can be described by a Lorentzian function, with $\Delta V_\perp /[1+ (\tau_s g \mu_B B_\perp / \hbar)^2 ]+V_{offset}$, where $g$ is the Landé g-factor of free holes in the Si, $\mu_B$ is the Bohr magneton, $\hbar$ is Planck's constant and $V_{offset}$ is the offset voltage[1, 28-31]. By taking $g = 2$, we extracted a $\tau_s$ of 101 ps for the sample A and 153 ps for the sample B at negative bias voltage, as indicated in Fig. 5 (a). For a positive bias voltage (Fig. 5 (b)), we also observed a similar trend with a shorter value of $\tau_s$ for the device with the polycrystalline MgO. This suggests that $\tau_s$ is affected by the crystalline quality of the tunnel contact, even though the same Si substrate was used.

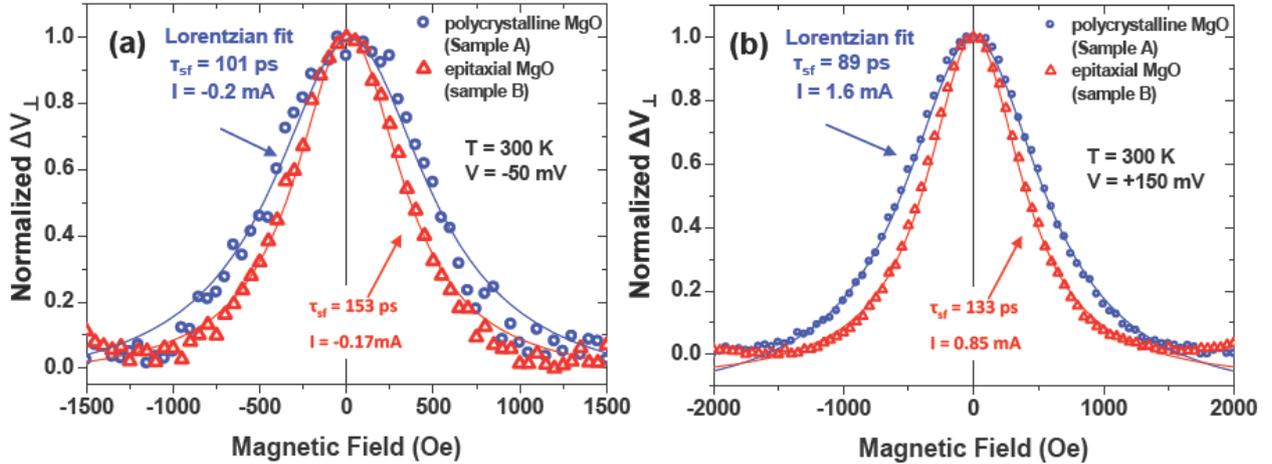

Figure 5: Hanle curves for the Fe/MgO/p-Si devices with a polycrystalline Fe/MgO (sample A, blue open circles) and an epitaxial Fe/MgO (sample B, red open triangles) at two different values of the bias voltage, as indicated. The hole spin lifetimes were extracted from fits to a Lorentzian function (blue and red lines).

Figure 6 (a) summarizes the extracted $\tau_s$ as a function of the bias voltage for polycrystalline (sample A) and epitaxial MgO (sample B). For the whole bias voltage region, the device with the polycrystalline MgO barrier exhibits shorter spin lifetime (95 ± 6 ps) than the device with the epitaxial MgO (143 ± 10 ps). Interestingly, these values are still larger than the value reported for the same p-Si with Fe/Al$_2$O$_3$ tunnel contacts[5], indicating that even polycrystalline MgO tunnel barrier is more effective for creating spins in Si than the amorphous Al$_2$O$_3$ tunnel barrier. It should be noted that the value of $\tau_s$ is affected by the $B_{loc}$, which is known to give rise to a broadening of the Hanle curve[5], and thereby an underestimation of the $\tau_s$. Therefore, the observed difference in $\tau_s$ is consistent with the results of RHEED and TEM observations, namely, the roughness at the Fe/MgO interface (and thus the magnitude of $B_{loc}$) for sample A is larger than that of sample B. We have also calculated the spin-diffusion length $l_{sf}$ which is given by $(D_h \tau_s)^{1/2}$, where $D_h$ is the diffusion coefficient of the holes. The hole mobility of 117 cm$^2$ V$^{-1}$ s$^{-1}$ measured by the Van der Pauw method gives the $D_h \approx 3.6$ cm$^2$ s$^{-1}$ at 300 K. Then, the calculated values of $l_{sf}$ are about 180 and 226 nm for the samples A and B, respectively. The obtained $l_{sf}$ are close to previous reports[1,5] and significantly larger than the channel length of state-of-the-art field-effect transistors (recall that $\tau_s$ and thus the calculated $l_{sf}$ is a lower bound). Therefore, this clearly demonstrates that communication of spin information in p-type Si is possible over the channel of transistors devices.

Figure 6 (b) shows the evolution of the spin-resistance-area product [spin-$RA \equiv (\Delta V_{spin}/I) \cdot A$] as a function of the bias voltage for the samples A and B. Two noticeable differences are observed : first, sample A exhibits an asymmetric behavior with respect to the bias polarity whereas the plot for sample *B* is nearly symmetric. The second remark concerns the magnitude of the spin-$RA$ that is higher for the epitaxial MgO in the complete bias voltage range. Particularly at low voltage (± 50 mV), the sample with the epitaxial MgO has a significantly higher value (about the double) than that of the sample with polycrystalline MgO. For the sake of completeness, we have measured the bias dependence of the resistance-area-product [Junction-$RA \equiv (V/I) \cdot A$], as shown in Fig. 6 (c). Both samples show very similar behavior, and this for the whole bias voltage range. The values of the junction-$RA$ estimated at low bias voltage are about 4 and 5 MΩµm$^2$ for samples A and B, respectively.

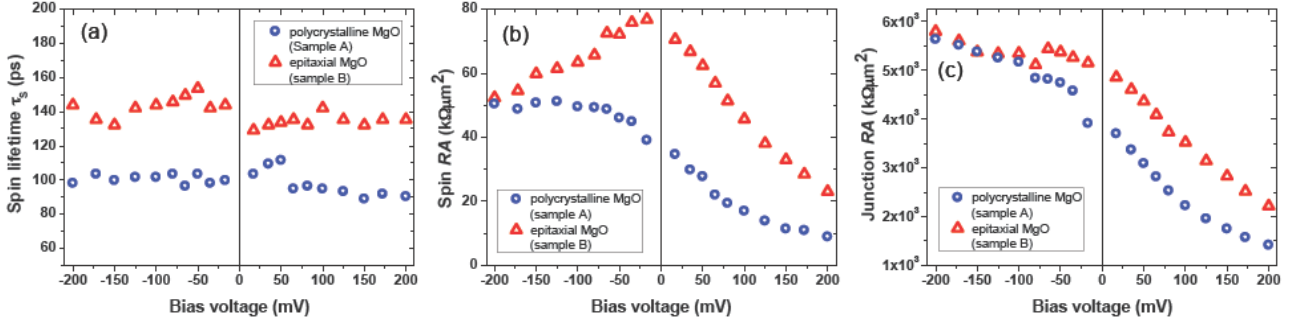

Figures 6 Bias dependence of the (a) spin lifetime $\tau_s$, (b) spin-$RA$ and (c) junction-$RA$ for the Fe/MgO/p-Si devices with polycrystalline MgO (sample A, blue open circles) and epitaxial MgO (sample B, red open triangles) barriers, respectively.

These experimental results suggest that a higher tunnel spin polarization is achieved in the epitaxial MgO barrier compared to that of the polycrystalline sample. This is consistent with a recent report showing that the crystalline quality of $AlO_x$ tunnel barrier can have an important effect on the efficiency of creating the spin accumulation[32]. Just as in a MgO-based MTJ[8-10], the spin filtering effect in tunnel contacts on Si is guaranteed by the 4-fold in-plane crystalline symmetry of the epitaxial MgO(001) barrier and its relationship with the bcc Fe electrode[33]. This leads to the conservation of the symmetry of the Bloch states at the Fermi level in the MgO layer, and as a consequence a higher tunnel spin polarization is observed.

Finally, the experimental value of the spin accumulation signal is compared to the value predicted from the standard theory for spin injection and spin diffusion in a nonmagnetic material predicted[34-36]. The measured spin-$RA$ at low bias voltages is about 45 k$\Omega\mu m^2$ for the sample A and is about the double (80 k$\Omega\mu m^2$) for the sample B, respectively (Fig. 6b). According to the standard theory, the spin-$RA$ is equal to $P^2 \cdot \rho_{Si} \cdot l_{sf}$ where $P$ is the tunnel spin polarization and $\rho_{Si}$ the resistivity of the substrate at 300 K (0.011 $\Omega$cm). Assuming $P = 0.6$ and $l_{sf} = 1$ µm, we can obtain the theoretical spin-$RA$ of 0.04 k$\Omega\mu m^2$. This is about 3 orders of magnitude smaller that the experimental values. Interestingly, several research groups using three-terminal Hanle measurements have also reported such large values in Si[1-7,30], Ge[11,12,14-17], and GaAs[29], no matter what type of tunnel barrier was used. As it was recently discussed[37,38], the origin of the discrepancies is not clear yet. Further experimental and theoretical work will be required to explain what governs the magnitude of the spin accumulation in semiconductors.

## 4. CONCLUSION

A detailed analysis of the growth conditions of the Fe/MgO tunnel contacts on p-type Si was reported. We found that the (2×1) Si surface is essential for achieving epitaxial MgO(001) on Si, whereas MgO grown on (1×1) shows polycrystalline structure. Clear Hanle and inverted Hanle effects were observed for polycrystalline and epitaxial Fe/MgO tunnel contacts by using a three-terminal configuration, demonstrating electrically induced spin polarization in p-Si at room temperature. We showed that the crystalline quality and the roughness of the tunnel barrier has a strong influence on the spin-related parameters. The device with epitaxial Fe/MgO contact exhibited longer spin lifetime and higher magnitude of the spin accumulation. The enhancement of the magnitude of the spin signal is most likely due to the spin filtering effect of the epitaxial MgO(001), which give rise to higher tunnel spin polarization of the tunnel contact. Our experimental findings demonstrate that a high quality MgO barrier is desirable to achieve efficient spin injection in Si.

## ACKNOWLEDGEMENTS

This work was supported by a JSPS Postdoctoral Fellowship for Foreign Researchers, the Funding Program for Next Generation World-Leading Researchers (No. GR099) and the Netherlands Foundation for Fundamental Research on Matter (FOM).